\documentclass[10pt,a4paper]{article}
\usepackage{epsfig}
\usepackage{amsmath}
\usepackage{amsfonts}
\usepackage{amssymb}
\usepackage{graphicx}

\begin{document}

\begin{flushright}
{\small LPTh-Ji 10/007}
\end{flushright}

\begin{center}
\vspace{2cm}

{\Large \textbf{\textit{A strong first order phase transition in the
UMSSM}}}

\vspace{1.3cm}

\textbf{Amine Ahriche}

\vspace{0.6cm}

\textit{Laboratory of Theoretical Physics, Department of Physics,
University of Jijel, PB 98 Ouled Aissa, DZ-18000 Jijel, Algeria.}

\textit{Email: ahriche@univ-jijel.dz}
\end{center}

\vspace{1.2cm}

\hrule \vspace{0.5cm} {\normalsize \textbf{{\large {Abstract}}}}

\vspace{0.5cm}

In this work, the electroweak phase transition (EWPT) strength has
been investigated within the $U(1)$ extended Minimal Supersymmetric
Standard Model (UMSSM) without introducing any exotic fields. We
found that the EWPT could be strongly first order for reasonable
values of the lightest Higgs and neutralino masses.

\vspace{0.8cm} \textbf{Keywords}: baryogenesis, electroweak phase
transition, extra singlet, extra gauge boson. \vspace{0.5cm}

\textbf{PACS}: 98.80.Cq, 11.10.Wx. \vspace{0.5cm} \hrule\vspace{2cm}

\section{Introduction}

The origin of matter-antimatter asymmetry is one of the main
problems of both particle physics and cosmology. The explanation of
this asymmetry $n_{b}/n_{\gamma}\sim10^{-10}$ \cite{wmap}, requires
the Sakharov criteria, which can be summarized in the existence of
such interactions in the early universe that violate the baryon
number $B$, the symmetries $C$ and $CP$ and occur out of
equilibrium. It appears that the Standard Model (SM) fulfills all
these criteria; the baryon number is not conserved at quantum level
due to the $B+L$ anomaly \cite{B+L}, a $CP$ violation source does
exist in $CKM$ matrix, and a departure from thermal equilibrium
could be reached through a strong first order phase transition
\cite{EB}, but its realization was not possible numerically due to
smallness of $CP$ violation effect and the weakness of the
electroweak phase transition (EWPT) \cite{RT}. In the SM, the EWPT
is so weak unless the Higgs mass is less than $45$ GeV
\cite{mhbound}, which is in conflict with present data \cite{LEP},
but it could be strong within some SM extensions (for e.g SM with a
singlet \cite{amin}).

In spite of its success and popularity, the MSSM with R-parity still
has two major problems: the $\mu$-problem \cite{mu} and the
potential proton decay problem due to dimension 5 operators
\cite{pdecay}. A natural solution to these problems would probably
require that the MSSM be extended by a new mechanism or a new
symmetry. The $U(1)^{\prime}$-extended MSSM (UMSSM) \cite{UMSSM} is
a straightforward extension of the MSSM with a non-anomalous TeV
scale Abelian gauge symmetry. In the minimal supersymmetric standard
model (MSSM), the EWPT could be strongly first order if the light
stop is lighter than the top quark \cite{MSSMPT}, and in the UMSSM,
it is also strongly first order but with the price of introducing 3
new extra singlet scalars \cite{lang}, or by adding new extra heavy
singlet fermions \cite{ham}. In this work, we will investigate the
possibility of getting a strong first order phase transition within
the minimal gauge extension of the MSSM, UMSSM without adding any
new field beside the usual singlet.

This paper is organized as follow: in the second section, we give a brief
review of the UMSSM model, define the scalar potential and discuss different
constraints on the parameters. After that, we discuss the EWPT dynamics and
show how to get a first order phase transition, and discuss our numerical
results. Finally, we summarize our results.

\section{The UMSSM model}

In the $U(1)^{\prime}$MSSM (or UMSSM), the gauge group is extended to
$G=SU(3)_{c}\times SU(2)_{L}\times U(1)_{Y}\times U(1)^{\prime}$ with the
couplings $g_{3}$, $g_{2}$, $g_{1}$ and $g^{\prime}$, respectively
\cite{UMSSM}. The superpotential is given by
\begin{equation}
W=\lambda S\epsilon_{ij}H_{1}^{i}H_{2}^{j}+Y_{U}\epsilon_{ij}Q^{i}U^{c}%
H_{2}^{j}+Y_{D}\epsilon_{ij}Q^{i}D^{c}H_{1}^{j}+Y_{L}\epsilon_{ij}L^{i}%
E^{c}H_{1}^{j},
\end{equation}
where $\epsilon_{ij}$ is the anti-symmetry $2\times2$ tensor, $Y_{U}$, $Y_{D}$
and $Y_{L}$ are Yukawa couplings, and $\lambda$\ is a coupling constant in
which $\lambda\left\langle S\right\rangle $ replaces the $\mu$-term in the
MSSM. The particle content, here, is the same as the MSSM in addition to a new
singlet and gauge boson; and the superpartners.

\subsection{The scalar potential}

In case where both sneutrinos or/and squarks do not develop vevs, the scalar
potential is the combination of the so-called $D$, $F$ and soft terms, which
are given by
\begin{align}
V_{D} &  =\frac{g_{2}^{2}+g_{1}^{2}}{8}\left(  H_{2}^{+}H_{2}-H_{1}^{+}%
H_{1}\right)  ^{2}+\frac{g_{2}^{2}}{2}\left\vert H_{1}^{+}H_{2}\right\vert
^{2}+\frac{g^{\prime2}}{2}\left\vert Q_{1}H_{1}^{+}H_{1}+Q_{2}H_{2}^{+}%
H_{2}+Q_{S}\left\vert S\right\vert ^{2}\right\vert ^{2},\nonumber\\
V_{F} &  =\left\vert \lambda\right\vert ^{2}\left\{  \left\vert \epsilon
_{ij}H_{1}^{i}H_{2}^{j}\right\vert ^{2}+\left\vert S\right\vert ^{2}\left[
H_{1}^{+}H_{1}+H_{2}^{+}H_{2}\right]  \right\}  \nonumber\\
V_{soft} &  =m_{H_{1}}^{2}H_{1}^{+}H_{1}+m_{H_{2}}^{2}H_{2}^{+}H_{2}+m_{S}%
^{2}\left\vert S\right\vert ^{2}+\left\{  A_{\lambda}S\epsilon_{ij}H_{1}%
^{i}H_{2}^{j}+h.c\right\}  .\label{Vv}%
\end{align}
Here $m_{H_{1}}^{2}$,~$m_{H_{2}}^{2}$,~$m_{S}^{2}$ and
$A_{\lambda}$\ are usually called soft parameters.\ The charges
$Q^{\prime}$s should be chosen in a way that ensures the anomaly
cancelations. The structure of the tree-level potential seems to
allow two explicit $CP$ violating relative phases between the scalar
vevs, but these phases could be canceled by such gauge
transformation, and the ground state is independent of any $CP$
violations phases. However, these relative phases could appear at
one loop through the superpartner masses (see Appendix A in
\cite{AN}).

In this setup, the scalar potential (\ref{Vv}) may admit another minimum
$\left(  0,0,x\right)  $, This wrong vacuum could play a very important role
in the EWPT dynamics.

\subsection{The parameters}

In this model, we have many parameters, some of them are free
like:\ $g^{\prime}$, $\lambda$, $\upsilon_{x}$, $\tan\beta=\upsilon
_{1}/\upsilon_{2}$, and the soft terms: $m_{Q}$, $m_{U}$, $A_{t}$,
$A_{\lambda}$, $M_{2}$, $M_{1}$ and $M_{1}^{\prime}$; and others are fixed by
such measured physical quantities:\ $g_{1}$, $g_{2}$, $\upsilon$ and $y_{t}$;
or such conditions like the elimination $m_{1}^{2}$, $m_{2}^{2}$ and
$m_{S}^{2}$ by imposing ($\upsilon_{1},\upsilon_{2},\upsilon_{x}$) to be the
absolute minimum of the effective potential at zero temperature. In addition
to the constrains that are coming from the extra $U(1)^{\prime}$ gauge
interactions, like mixing between the neutral gauge boson $Z$ and the new one
$Z^{\prime}$ \cite{MMz}
\begin{equation}
2M_{ZZ^{\prime}}^{2}/(M_{Z^{\prime}Z^{\prime}}^{2}-M_{ZZ}^{2})<10^{-3},
\label{zmix}%
\end{equation}
and the bound on the heavy $Z^{\prime}$ mass \cite{Zplimit}
\begin{equation}
M_{Z^{\prime}}>(500-800)~GeV. \label{zpm}%
\end{equation}

Keeping the new charges as free parameters beside the condition $Q_{1}%
+Q_{2}+Q_{s}=0$, we take random values for the free parameters, taking into
account such conditions like the elimination of $m^{2}$'s parameters in
(\ref{Vv}), the perturbativity of the quartic couplings in (\ref{Vv}), and the
vacuum stability. Another constrain could be derived from the upper bound on
the mixing between the gauge boson $Z$ and the new one $Z^{\prime}$
(\ref{zmix}), and the lower bound on the new gauge boson $Z^{\prime}$ mass
(\ref{zpm}). The condition (\ref{zpm}), could be achieved by considering
relatively large $\upsilon_{x}$, or large $g^{\prime}Q^{\prime}$s. The
condition (\ref{zmix}) could be fulfilled by vanishing the mixing term
$M_{ZZ^{\prime}}^{2}$, i.e., if%
\begin{equation}
Q_{1}=Q_{2}\tan^{2}\beta, \label{zmixing}%
\end{equation}
or if $M_{Z^{\prime}Z^{\prime}}^{2}>>M_{ZZ}^{2},M_{ZZ^{\prime}}^{2}\,$, which
roughly means
\begin{equation}
g^{\prime}\left\vert Q_{S}\right\vert {\upsilon_{x}}\gtrsim(500-800)~GeV,
\label{supmixing}%
\end{equation}
or a serious tuning in the values of $Q_{1,2}$ and $tan\beta$.

In our search for the parameter's space that fulfills the strong first-order
phase transition criterion: $\upsilon\left(  T_{c}\right)  /T_{c}>1$, we will
focus our search on two regions:

\textbf{a)} Moderate values for the parameters $Q_{1,2}$ and $tan\beta$, where
(\ref{zmix}) is nearly satisfied. In this case, the singlet vev $\upsilon_{x}%
$, could be the order $\upsilon$ or even relatively smaller.

\textbf{b)} The two terms $M_{ZZ}^{2}$ and $M_{ZZ^{\prime}}^{2}$ in the
mass-squared matrix of $(Z,Z^{\prime})$, should be suppressed with respect to
the mass term $M_{Z^{\prime}Z^{\prime}}^{2}$. In this case, the values of
$U^{\prime}(1)$ charge and the vev of the singlet, $Q_{s}$ and $\upsilon_{x}$,
should be large enough (\ref{supmixing}).

The mass parameters:\ $m_{Q}$, $m_{U}$, $A_{t}$, $M_{2}$, $M_{1}$ and
$M_{1}^{\prime}$ appear at one-loop level in the effective potential,
therefore we expect that their role is less important in the EWPT dynamics.
But each of these parameters:\ $g^{\prime}$, $\lambda$, $\upsilon_{x}$ and
$\tan\beta$, as well the charges $Q^{\prime}$s that appear multiplied by
$g^{\prime}$, seems to be very important, therefore we focus on these
parameters while fixing the mass parameters: $g^{\prime}=g_{1},~m_{Q}%
=m_{U}=1~TeV,$ and giving different values for the rest $A_{t},~M_{2},~M_{1}$
and $M_{1}^{\prime}$.

\section{The electroweak phase transition}

\subsection{The phase dynamics}

Due the condition (\ref{zmix}) and (\ref{zpm}), there could exist a hierarchy
between the vev of the singlet and those of the doublets, i.e., $\upsilon
_{x}>>\upsilon_{1,2}$. In this case, the gauge symmetry could be broken in two
step: $(0,0,0)\overset{T_{c}^{\prime}}{\rightarrow}(0,0,x)\overset{T_{c}%
}{\rightarrow}(\upsilon_{1},\upsilon_{2},\upsilon_{x})$, or just in once
$(0,0,0)\overset{T_{c}}{\rightarrow}(\upsilon_{1},\upsilon_{2},\upsilon_{x})$.
Since the singlet dynamics does not affect the $SU(2)$ sphaleron processes, we
will not be interested in distinguishing between the one- and two-steps
symmetry breaking. We will treat our field dynamics using the effective
potential where the singlet is replaced by it thermal vev. At the critical
temperature $T_{c}$, the two minima get degenerate%
\begin{equation}
V_{eff}\left(  \upsilon_{1}^{c},\upsilon_{2}^{c},\upsilon_{x}^{c}%
,T_{c}\right)  =V_{eff}\left(  0,0,x_{c},T_{c}\right)  .
\end{equation}
Below this temperature, the new minimum becomes the absolute one, and the
system has to move from the old (false) vacuum to the new (true) one. In the
case where a barrier does exist between the two minima, this transition has to
occur via tunneling trough bubbles nucleation at certain points, which expand
and \ fill the whole space by the new vacuum $\upsilon_{1,2}\left(  T\right)
\neq0$, i.e, the symmetry is broken.

The $B+L$ anomalous interactions \cite{B+L}, that violate the baryon number
have not the same rate in the symmetric and broken phases (i.e., at both sides
of the bubble wall). In the symmetric phase, this rate behaves like $\sim
T^{4}$ \cite{Symm}, and suppressed as $exp(-E_{Sp}/T)$ \cite{bro}, in the
broken phase, where $E_{Sp}$ is the system static energy within such field
configuration called the sphaleron \cite{Sph}. Therefore any generated baryon
number at the symmetric phase will erase at the broken phase, unless these
interactions are switched off at the broken phase, this is translated to the
famous criterion \cite{SFOPT}.

Here, we are interested in the in the value of the EW vev at the transition
temperature, and since the singlet field does not play an important role in
the sphaleron processes \cite{amin}, the criterion of a strong first order
phase transition in our case is given by \cite{SFOPT}%
\begin{equation}
\upsilon\left(  T_{c}\right)  /T_{c}\equiv\sqrt{\upsilon_{1}^{2}\left(
T_{c}\right)  +\upsilon_{2}^{2}\left(  T_{c}\right)  }/T_{c}>1.\label{cri}%
\end{equation}

In the general case where the relative phases $\theta_{1,2}\neq0$,
the field ground state should be written as $\left\{  v_{i}\right\}
_{i=1,5}$= ($\upsilon_{1}$, $\upsilon_{2}\cos\theta_{1}$,
$\upsilon_{2}\sin\theta_{1}$, $\upsilon_{x}\cos\theta_{2}$,
$\upsilon_{x}\sin\theta_{2}$)\ instead of
($\upsilon_{1},\upsilon_{2},\upsilon_{x}$) and the two relative
phases. These 5 variables should be treated independently when
looking for $\upsilon _{1,2,x}$\ and $\theta_{1,2}$\ at any
temperature $T$. Then the phase
transition could be defined through the equations%
\begin{equation}
\frac{\partial}{\partial v_{i}}V_{eff}\left(  v_{i},T_{c}\right)
=0,~V_{eff}\left(  v_{1},v_{2},v_{3},v_{4},v_{5},T_{c}\right)  =V_{eff}\left(
0,0,0,x_{1},x_{2},T_{c}\right)  ,\label{TC}%
\end{equation}
where $x_{1,2}$\ are the real and imaginary parts of the singlet vev in the
wrong vacuum.

\subsection{Numerical results}

\begin{figure}[h]
%\begin{centre}
\includegraphics[width=8cm,height=6cm]{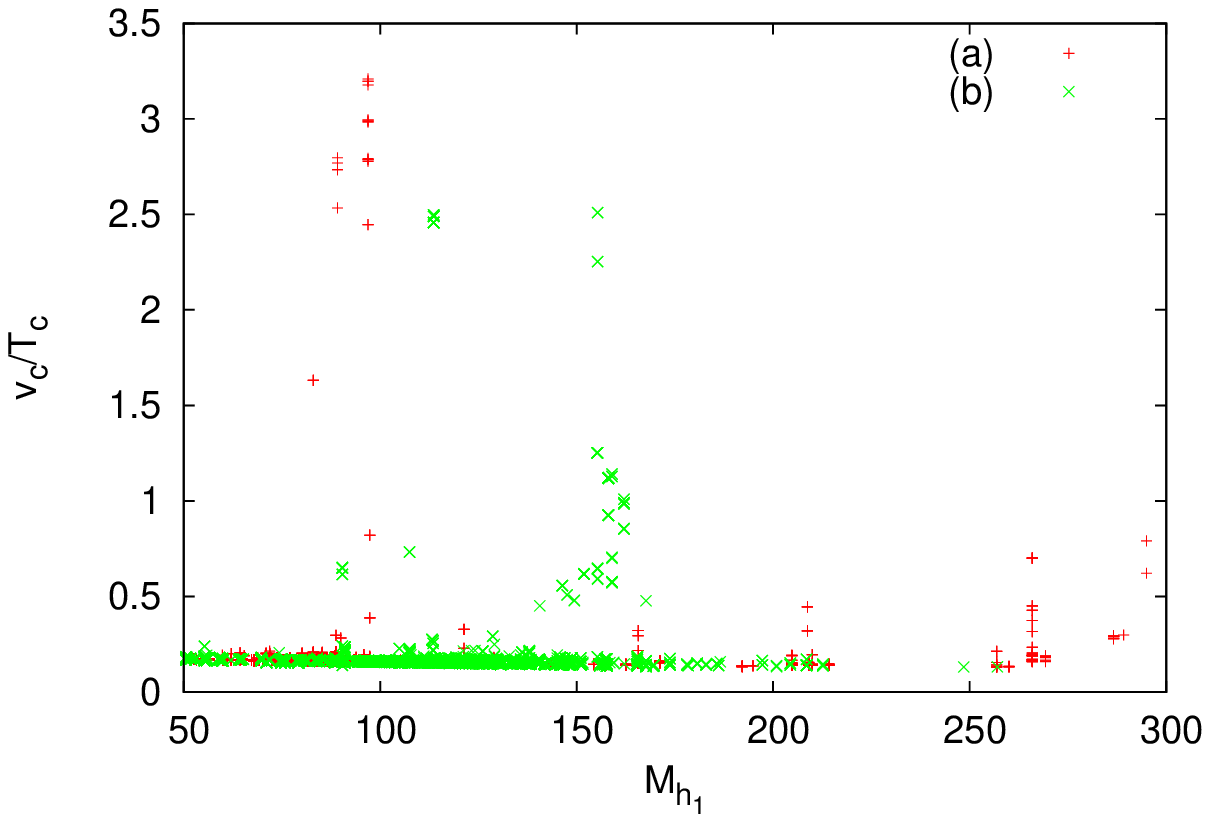}~~\includegraphics[width=8cm,height=6cm]{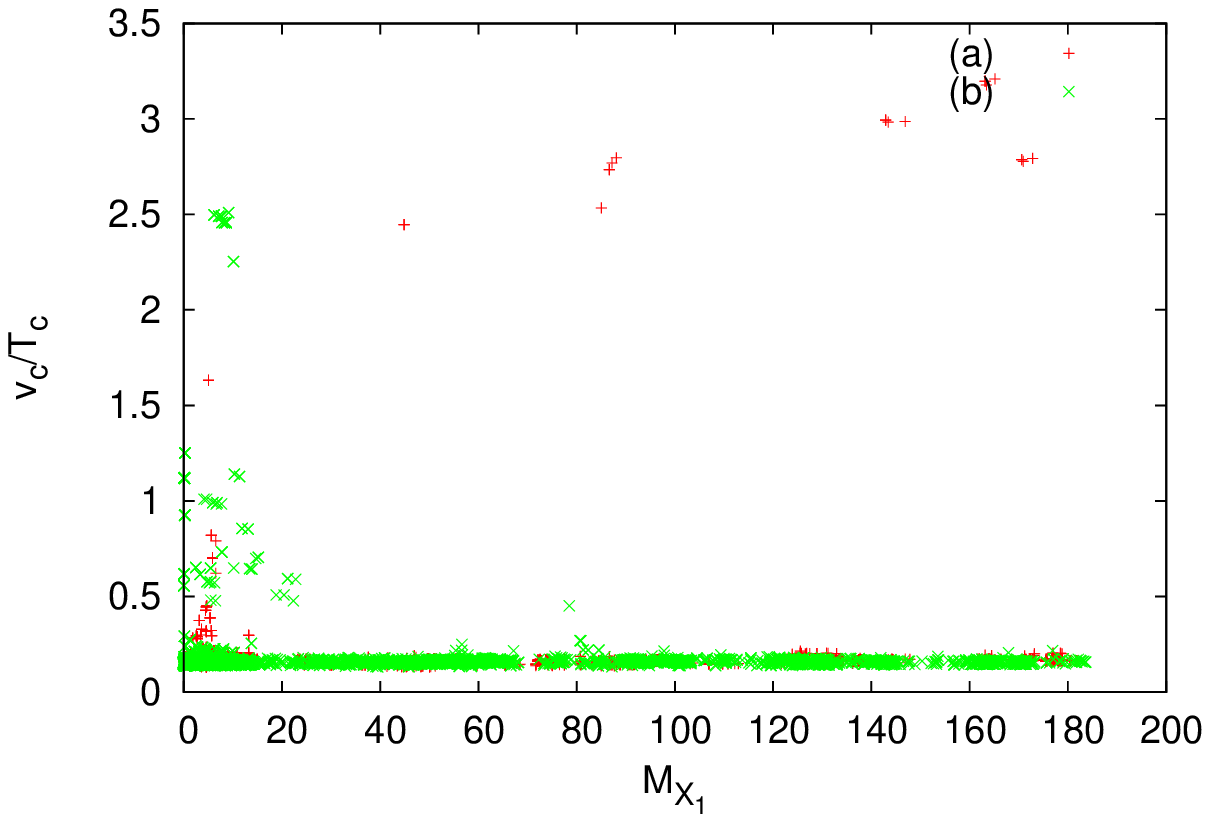}
%\end{centre}
\caption{\textit{The dependence of the quantity
$\upsilon_{c}/T_{c}$ on the lightest Higgs (right) and the lightest
neutralino (left) masses. The green points refer to cases in region
\textbf{(a)}, and the red ones to cases in
region \textbf{(b).}}}%
\label{Mhn}%
\end{figure}

In the following figures, we show the dependence of the quantity
$\upsilon\left(  T_{c}\right)  /T_{c}$ in (\ref{cri}) on the
lightest Higgs mass (in Fig. \ref{Mhn} left), and on the lightest
neutralino mass (in Fig. \ref{Mhn} right), for a random choice of
about 2000 cases in both regions \textbf{(a)} and \textbf{(b)},
where different conditions are fulfilled.

As it is clear from Fig. \ref{Mhn}, the EWPT could be strongly first
order in the two regions (a) and (b), for different values of the
lightest Higgs mass. The fact that the lightest Higgs mass is around
90$\sim$100 GeV might be consistent with experiment because the
doublet couplings will be modified due to the mixing with singlet.
The EWPT could be also, strongly first order for different the
lightest neutralino masses. This leads to the conclusion that the
EWPT strength is a factor beyond the fields contributions to the
effective potential at zero temperature.

The critical temperature is, in general, higher when comparing with
minimal SM ($\sim$100 $GeV$), the generic value is larger than $500$
$GeV$, this is a consequence of the the interaction of the doublets
with the singlet that has, in general, a very large vev. However for
the benchmarks, giving a strong first order EWPT, it is relatively
smaller than the generic values. As it is small, the EWPT is
stronger. In order to understand this point, we take a benchmark
from Fig. \ref{Mhn}, and study the dependance of the scalar vevs on
the temperature $T$. We will see also how could this behavior be
changed with respect to the charges $Q^{\prime}$s, and some
parameters like $A_{t}$ and $\theta_{1,2}^{(v)}$, that appear in the
effective potential at one-loop level. Therefore we consider the
following modifications:

\begin{table}[h]%
\begin{tabular}
[c]{c|c|c|c|c|c|c|c|c|c|}\cline{2-10}
& $Q_{1}$ & $Q_{2}$ & $A_{t}$ & $\theta_{1}^{(v)}$ & $\theta_{2}^{(v)}$ &
$T_{c}$ & $\upsilon_{c}/T_{c}$ & $m_{h_{1}}$ & $m_{\chi_{1}}$\\\hline
\multicolumn{1}{|c|}{(1)} & $20/7$ & $-4/7$ & $300$ & $0$ & $0$ & $73.25$ &
$3.454$ & $116.34$ & $7.70$\\\hline
\multicolumn{1}{|c|}{(2)} & $20/7$ & $-4/7$ & $300$ & $\pi/9$ & $-\pi/9$ &
$75.72$ & $3.338$ & $116.34$ & $8.32$\\\hline
\multicolumn{1}{|c|}{(3)} & $2$ & $1$ & $300$ & $0$ & $0$ & $629.17$ & $0.149
$ & $112.31$ & $7.70$\\\hline
\multicolumn{1}{|c|}{(4)} & $20/7$ & $-4/7$ & $1500$ & $0$ & $0$ & $274.13$ &
$0.198$ & $116.34$ & $7.70$\\\hline
\end{tabular}
\caption{\textit{Values of the parameters used to study the scalar vevs
dependance with respect to the temperature. We used common values for the
parameters: }$\lambda=0.01$\textit{, }$tan\beta=1$\textit{, }$A_{\lambda
}=1210~GeV$\textit{, }$\upsilon_{x}=1046.4~GeV$\textit{, }$M_{1}%
=M_{2}=100~GeV$\textit{, }$M_{1}^{\prime}=300~GeV$\textit{. The mass-dimension
parameters are given in GeV.}}%
\label{Tab}%
\end{table}
From this table, it is clear that the EWPT strength is sensitive to
the relative phases $\theta_{1,2}^{(v)}$, that appear in the
effective potential at one-loop through the superpartners masses.
For this parameters set, these relative phase shift the critical
temperature by some units. The effect of the parameter $A_{t}$, that
appears in the effective potential through the one-loop corrections
of the stops, is more important. The critical temperature, in this
case (4) is shifted by 200 GeV, which makes the EWPT so weak. From
case (3), the effect of the charges $Q^{\prime}$s is extremely
important.

\begin{figure}[h]
\includegraphics[width=16cm,height=6cm]{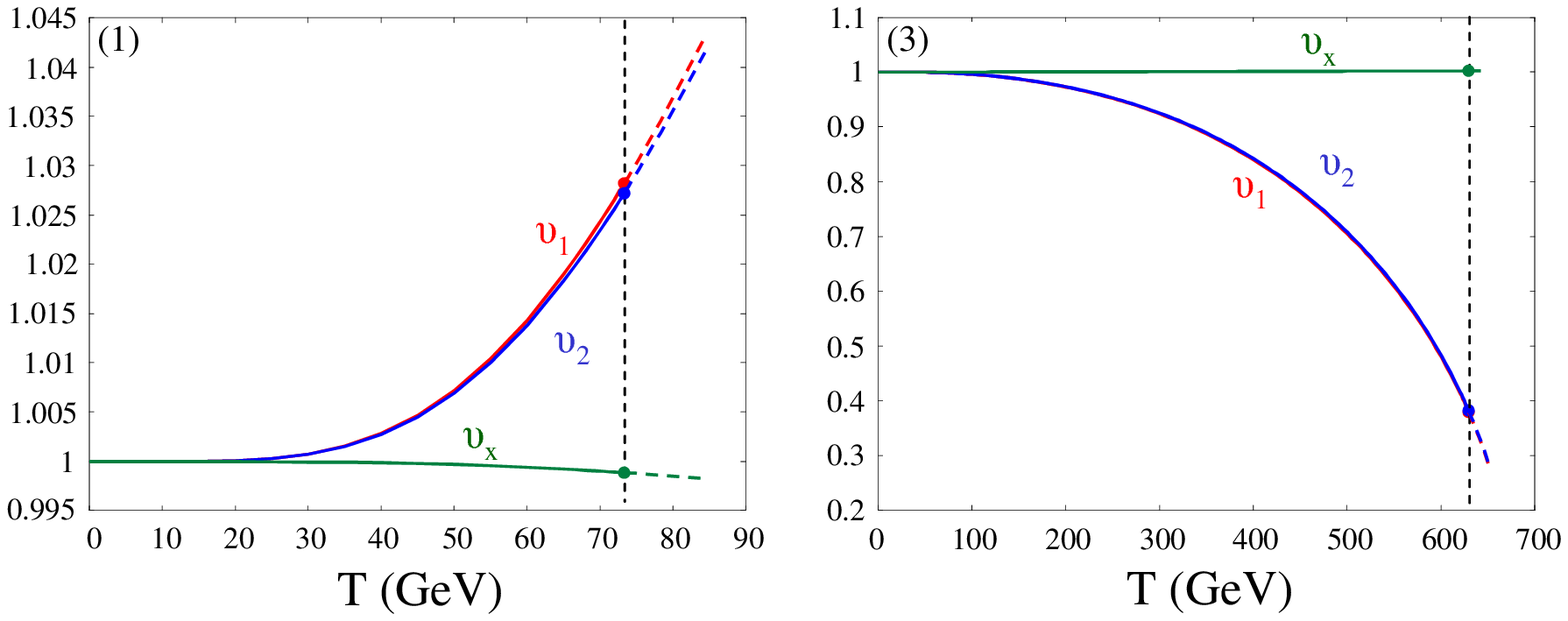}\caption{\textit{The
dependence of the scalar vevs on the temperature T. These quantities are
scaled by their zero temperature values. The solid curves refer to the broken
phase while the dashed ones refer to the symmetric phase, where (0,0,x(T)) is
still the global minimum.}}%
\label{cs}%
\end{figure}

In Fig. \ref{cs}, we show the dependance of the ground state on the
temperature below the critical temperature for the benchmark (1) in
Table-\ref{Tab}, and its modifications (2)-(4). From Fig. \ref{cs}, one
remarks that the common feature between all these cases is that the dependance
of the singlet vev on the temperature is very weak around and below the
critical temperature.

Another important remark, is that the Higgs vevs for this benchmark
(Fig. \ref{cs}-1), are increasing when the Universe gets cooled
unlike the SM, or MSSM. In case (3) (Fig. \ref{cs}-3), the Higgs
vevs decrease, but slower than the SM or MSSM. This can be seen
clearly in Fig. \ref{Mhn}-3, for the benchmarks whose critical
temperature is between $250\sim400$ $GeV$, where the corresponding
value of $\upsilon\left( T_{c}\right) /T_{c}$ is larger when
comparing with the SM. The fact that the doublets vev values at high
temperatures are larger than their zero temperature values has been
mentioned in a similar work \cite{ham}. This behavior: increasing
Higgs vevs w.r.t the temperature or decreasing slower than the SM
case, is a consequence of the interaction of the singlet with the
doublet. These interactions lead to relax the shape of the potential
in the direction of the doublets, and therefore enhances the ratio
in (\ref{cri}), and strengthen the EWPT. This is a common feature
for models with singlets \cite{amin}.

Unlike the SM and MSSM, the wrong vacuum $(0,0,x(T))$, is evolving
w.r.t the temperature, therefore its evolution could be a very
important factor that strengthens the EWPT. It could delay the
transition, and then enhances the ratio (\ref{cri}). The EWPT
dynamics seems to be less sensitive to some input parameters like
$m_{Q},~m_{U},$\ $A_{t},~M_{2},~M_{1}$, $M_{1}^{\prime}$ and
especially the relative phases $\theta_{1,2}^{(v)}$, since they
appear in the effective potential at one loop, however, we have
shown in Table-\ref{Tab} that the EWPT dynamics is sensitive to
these parameters more than expected. This sensitivity is a
consequence of the dependance of the two vacua (instead one) on
these input parameters.

The possible one-loop spontaneous $CP$ violation due to the relative
phases $\theta_{1,2}^{(v)}$, could have an important
phenomenological impact. An important feature in this model is that
the dark matter candidate (lightest neutralino) could be dominated
by the new ingredient $\tilde{B}^{\prime}$. This possibility will be
investigated in further work \cite{next}.

\section{Conclusion}

In this work, the electroweak phase transition nature within the
minimal $U(1)$ extension of the MSSM (UMSSM) has been investigated.
We found that the EWPT could be strongly first order for reasonable
masses of scalar Higgs bosons and the lightest neutralino, without
adding extra singlet scalars or fermions. We evaluated the effective
potential at one-loop taking into account the whole particle
spectrum, and its temperature dependant corrections were estimated
exactly using the known techniques. We found that EWPT strength
could be enhanced due to two factors: first, the interactions of the
singlet scalars with the doublets that relax the shape of the
effective potential in the doublets directions, which lead to a
large value for the ratio $\upsilon\left(  T_{c}\right)  /T_{c}$, at
the critical temperature. The second factor is that the
temperature-dependant local minimum $(0,0,x\left(  T\right)  )$,
could play an important role during the EWPT dynamics. It can delay
the phase transition until relatively low temperatures (even below
$100$ $GeV$), which favor the ratio $\upsilon\left(  T_{c}\right)
/T_{c}$ to be large enough.

We mention also that the reliability of the parameter choice, as well as the
EWPT strength are more sensitive to the input parameters that appear in the
effective potential at one-loop.

\section*{Acknowledgments} This work was supported by the Algerian
Ministry of Higher Education and Scientific Research under the
cnepru-project D01720090023.

\end{document}